\documentclass[tighten,twocolumn]{aastex631}
\usepackage{graphicx}
\usepackage{ulem}
\usepackage{amsmath}
\usepackage{wrapfig}
\usepackage[thinlines]{easytable}
\usepackage{tikz}
\usepackage{hyperref}
\usepackage{ulem}

\usepackage{float}
\usepackage{longtable}
\usepackage{booktabs}
\usepackage[percent]{overpic}

\begin{document}

\shortauthors{K{\i}ro\u{g}lu et al.}

\title{Spin-Orbit Alignment in Merging Binary Black Holes Following Collisions with Massive Stars}

\correspondingauthor{Fulya K{\i}ro\u{g}lu}
\email{fulyakiroglu2024@u.northwestern.edu}

\author[0000-0003-4412-2176]{Fulya K{\i}ro\u{g}lu}
\affiliation{Center for Interdisciplinary Exploration \& Research in Astrophysics (CIERA) and Department of Physics \& Astronomy \\ Northwestern University, Evanston, IL 60208, USA}

\author[0000-0002-7444-7599]{James C.\ Lombardi Jr.}
\affiliation{Department of Physics, Allegheny College, Meadville, Pennsylvania 16335, USA}

\author[0000-0002-4086-3180]{Kyle Kremer}
\affiliation{Department of Astronomy \& Astrophysics, University of California, San Diego; La Jolla, CA 92093, USA}

\author[0009-0009-6575-2207]{Hans D.\ Vanderzyden}
\affiliation{Department of Physics, Allegheny College, Meadville, Pennsylvania 16335, USA}

 \author[0000-0002-7132-418X]{Frederic A. Rasio}
\affiliation{Center for Interdisciplinary Exploration \& Research in Astrophysics (CIERA) and Department of Physics \& Astronomy \\ Northwestern University, Evanston, IL 60208, USA}

\begin{abstract}

Merging binary black holes (BBHs) formed dynamically in dense star clusters are expected to have uncorrelated spin--orbit orientations
since they are assembled through many random interactions. However, measured effective spins in BBHs detected by LIGO/Virgo/KAGRA  hint at additional physical processes that may introduce anisotropy. Here we address this question by exploring the impact of stellar collisions, and accretion of collision debris, on the spin--orbit alignment in merging BBHs formed in dense star clusters. 
Through hydrodynamic simulations, we study the regime where the disruption of a massive star by a BBH causes the stellar debris to form individual accretion disks bound to each black hole.
We show that these disks, which are randomly oriented relative to the binary orbital plane after the initial disruption of the star, can be reoriented by strong tidal torques in the binary near pericenter passages. Following accretion by the BHs on longer timescales, BBHs with small but preferentially positive effective spin parameters ($\chi_{\rm eff} \lesssim 0.2$) are formed. Our results indicate that BBH collisions in young massive star clusters could contribute to the observed trend toward small positive $\chi_{\rm eff}$, and we suggest that the standard assumption often made that dynamically assembled BBHs should have isotropically distributed BH spins is not always justified. 

\end{abstract}
 
\section{Introduction}

 Observational breakthroughs over the past decade have provided compelling evidence for the presence of black holes (BHs) in globular clusters (GCs). X-ray and radio detections \citep{Strader2012}, along with radial velocity measurements \citep{Giesers2018}, have confirmed that BHs can remain bound to GCs, contrary to previous theoretical predictions that dynamical interactions would eject them early in the cluster's evolution. Complementing these observations, advanced N-body simulations \citep[e.g.,][]{Morscher_2015,Wang2016_DRAGON,Askar2018,Kremer2020_catalog} have demonstrated that BHs influence the long-term dynamics and structure of GCs, particularly in their dense cores. It is now widely accepted that most old GCs harbor \textit{dozens to hundreds} of dynamically active BHs today, playing a pivotal role in delaying gravothermal collapse and shaping the cluster's overall evolution \citep{Merritt2004,Mackey2007,BreenHeggie2013,Kremer_2018_ngc3201,Kremer2019,Kremer2020_bhburning}.

Additionally, recent gravitational wave (GW) detections by LIGO/Virgo/KAGRA (LVK) have provided groundbreaking evidence for the formation and merger of stellar-mass binary BHs \citep[BBHs;][]{Abbott_2023}. Although the formation pathways of these mergers remain an active area of research \citep{MandelBroekgaarden2022}, dynamical formation in dense stellar systems, such as GCs, has emerged as a likely key contributor \citep[e.g.,][]{Rodriguez2016}, competing with traditional isolated binary evolution scenarios \citep[e.g.,][]{Belczynski_2016}. These dense environments provide the necessary conditions to drive frequent close encounters, enabling the formation, hardening, and eventual merger of BBHs.

Various recent studies have highlighted several key properties of the BBH population expected to form in clusters, including their masses \citep[e.g.,][]{Rodriguez2019_TNG,Kremer2020_imbh}, eccentricities \citep[e.g.,][]{Zevin2021}, and spins \citep[e.g.,][]{Payne2024}. If observed, these features may provide crucial tests of the role of clusters in forming GW sources. Regarding spins in particular, the spin magnitude and orientation of BBH components are determined by the angular momentum accreted from their progenitors and by their interactions with nearby astrophysical objects. Canonically, random interactions in dense environments have been linked to uncorrelated spin-orbit orientations and an isotropic distribution in effective spin \citep[e.g.,][]{Rodriguez2016_spins}, which is defined by
\begin{equation}
  \chi_{\rm eff} = \frac{M_1\chi_1 \cos{\theta_1} + M_2\chi_2 \cos{\theta_2}}{M_1+M_2} \in (-1,1) \label{eq:chi_eff}
\end{equation}
where 
$M_i$ is the BH mass, and $\chi_i\equiv S_i/M_i^2 \in (0,1)$ denotes the dimensionless spin of each BH, with $S_i $ representing the spin angular momentum of the BH, and $\theta_i$ is the angle between the vector ${S}_i$ and the orbital angular momentum.
However, evidence suggesting a preference against a purely isotropic spin distribution \citep{Abbott_2023} poses a challenge to the hypothesis that BBHs form predominantly in dense star clusters through random dynamical encounters \citep[e.g.,][]{Rodriguez_2021}.

Recent studies reveal that BHs in clusters often undergo close encounters with stars, leading to unique electromagnetic (EM) transients and, in some cases, significantly altering BH properties. These encounters may cause a star to pass by a BH within its tidal disruption radius or to collide physically with the BH, resulting in the complete or partial disruption of the star \citep[e.g.,][]{Perets_2016,Lopez_2019,Kremer2019_tde,Ryu_2020b,Kremer_2022,Kremer2023_tde,Kiroglu_2023}. Our recent work demonstrates collisions with massive stars, comparable in mass to the BHs, can result in substantial accretion and spin-up, with important implications for GW detections by LVK \citep{Kiroglu_2024}.

$N$-body simulations show that the majority of BH-star collisions occur during multi-body encounters \citep{FregeauRasio2007,Kremer2021,Kiroglu_2024}. This trend is particularly pronounced in clusters with a high initial binary fraction among massive stars \citep[e.g.,][]{Gonzalez_2021}, consistent with observational data indicating that nearly all O- and B-type stars in the Galactic field are born in binaries \citep{Sana_2012}. Motivated by these trends, in this paper we conduct hydrodynamic simulations to study encounters between BBHs and massive stars. We explore a portion of the parameter space for BBH properties (e.g., semi-major axis, eccentricity, and component masses) guided by $N$-body simulations of dense star clusters.

\begin{figure*}
    \centering
    \includegraphics[width=0.8\linewidth]{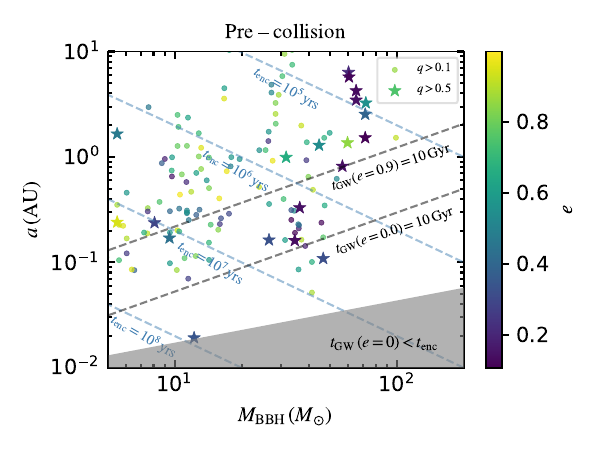}
    \caption{\footnotesize Comparison of the GW inspiral timescale (black dashed lines) of a BBH with semi-major axis $a$ and total mass $M_{\rm BBH}$ (assuming equal-mass components) to the dynamical encounter timescale (blue dashed lines) in a typical dense star cluster with a central density of $n\sim 10^6\,\rm{pc}^{-3}$ and velocity dispersion $10\,{\rm km \, s^{-1}}$. The shaded gray region displays the parameter space where $t_{\rm GW}<t_{\rm enc}$, indicating that BBHs merge via GW emission before they have a chance to dynamically interact with a star. Above this region, the points denote all BBH-star collisions identified in the cluster models of \citet{Kiroglu_2024}, with colors representing their orbital eccentricity at the time of the collision. Different symbols indicate collisions between BHs and stars of different mass ratios ($q=M_{\star}/M_{\rm BH}$).
    }
        \label{fig:boundary}
\end{figure*}

Our focus is on BBHs that are initially wide enough to interact with stars before merging via GW emission, but also compact enough that both BHs can dynamically interact with the disrupted stellar debris. Figure~\ref{fig:boundary} sets the stage for our paper, where we demonstrate the parameter space for the BBH semi-major axis and total mass of interest. In gray we highlight the ``excluded'' region where the BBHs are likely to merge before undergoing a stellar encounter that may lead to a collision. We calculate the GW inspiral time $t_{\rm GW}$ of the binaries following \citet{Peters_1964} and the encounter timescale using $ t_{\rm enc}= ( n \Sigma_{\rm bs} \sigma_v)^{-1}$, where $n$ is the cluster's central number density and $\sigma_v$ is the central velocity dispersion, $\Sigma_{\rm bs}$ is the cross section for binary-single encounters. The scatter points represent all BBH-star collisions from our recent N-body simulations of dense star clusters \citep{Kiroglu_2024} using the \texttt{Cluster Monte Carlo} (\texttt{CMC}) code, a H\'{e}non-style $N$-body code for stellar dynamics \citep[see][for a detailed review]{Rodriguez_2022}. These simulations are performed with an initial total number of stars $N=8\times 10^5$, metallicity $Z=[0.1,1]\,Z_{\odot}$, initial virial radius $r_v = [0.5, 1]\,\rm{pc}$, and a primordial binary fraction for massive stars $f_b\,(>15\,M_{\odot})=100\%$. In these models, up to $20\%$ of all stellar disruptions by BHs occur during BBH+star three-body encounters, potentially enhancing the merger rates of these BBHs through orbital parameter modifications.  We find that approximately $60\%$ of these BBHs are dynamically assembled while the remaining $40\%$ originate from dynamically-shaped primordial binaries. 

Our paper is organized as follows. In Section \ref{sec:method}, we describe the computational methods used for hydrodynamic simulations and the calculation of BH spins resulting from stellar collisions and subsequent accretion. In Section \ref{sec:results}, we present the outcomes of the hydrodynamic simulations and discuss the varying results across the parameter space. Finally, we conclude and summarize our findings in Section \ref{sec:conc}.

\section{Methods}
\label{sec:method}

Hydrodynamic simulations of the BBH+star collision paramater space mapped out in Figure~\ref{fig:boundary} are the focus of this study. For this task, we use the \texttt{StarSmasher} code \citep{Rasio_1991,Gaburov_2018}, which employs smoothed particle hydrodynamics (SPH) to model stellar gas as particles with density profiles defined by a smoothing kernel. We adopt a Wendland C4 kernel \citep{Wendland_1995} with compact support for smoothing and gravitational softening. To prevent unphysical particle interpenetration, we implement artificial viscosity with a Balsara switch \citep{Balsara_1995} as described in \citet{Hwang_2015}. Gravitational forces are computed using direct summation on NVIDIA GPUs, which provides improved accuracy compared to tree-based methods \citep{Gaburov_2010b}. The simulations use an equation of state that incorporates contributions from both ideal gas and radiation pressure \citep{Lombardi_2006}. The SPH particles evolve using variational equations of motion \citep{Monaghan_2002,Springel_2002,Lombardi_2006}.

\subsection{Stellar Profiles}

We generate 1D stellar models using MESA \citep{Paxton_2011} for a $10\,M_{\odot}$ main-sequence (MS) star and a $9.6\,M_{\odot}$ post-MS star, assuming a metallicity of $Z = 0.1\,Z_{\odot}$.  The initial stellar parameters are summarized in Table \ref{table1}. We convert these 1D stellar models into 3D SPH initial conditions by interpolating density, pressure, and mean molecular weight profiles onto a hexagonal close-packed lattice, following the methods of \cite{Sills_2001,Hwang_2015}. Each simulation employs about $N\sim 10^5$ unequal-mass particles.
\begin{deluxetable}{lcccccc}
\tabletypesize{\footnotesize}
\label{table1}
\tablecolumns{7}
\tablewidth{10pt}
\tablehead{
    \colhead{star} & 
    \colhead{$M$} &
    \colhead{age} &
    \colhead{$R$} &
    \colhead{$\rho_c$} &
    \colhead{$T_{\rm eff}$} &
    \colhead{$N$} \\
    \colhead{} &
    \colhead{$(M_\odot)$} & 
    \colhead{(Myr)} & 
    \colhead{($R_{\odot}$)} &
    \colhead{$(\rm{g/cm^{3}})$} &
    \colhead{(K)} &
    \colhead{}
}
\startdata
10\,MS & 10 & $11$ & 3.95 & 12.5 & $2.7\times 10^4$ & 99,954\\ 
10\,post & 9.6 & $23$ & 8.05 & 713 & $2.2\times 10^4$ & 100,218\\ 
\hline
\enddata
\caption{The initial properties of all stars used in our SPH simulations. In Columns 2–7, we give star mass, age, radius, central density, effective temperature, and the number of SPH particles employed to create stellar profiles in our simulations, respectively. All models have metallicity $Z=0.1\,Z_{\odot}$.}
\end{deluxetable}

For the MS model, we use a constant number density of particles with $\sim 200$ neighbors each, and the desired density profile from MESA is recreated by appropriately setting the particle masses.  For the post-MS star, we vary the spacing of particles using a stretchy HCP lattice \citep{Gibson_2024} with $\alpha=0.15$ and $\sim 600$ neighbors. After initialization, we allow the SPH fluid to settle into hydrostatic equilibrium by applying an artificial drag force to damp oscillations for a
few hundred dynamical timescales
as described in \citet{Lombardi_2006}. We run all the simulations until the star is fully disrupted and the simulation reaches a slowly varying state. Figure \ref{fig:mesa} in the Appendix compares the relaxed SPH density and mass profiles with those from MESA, showing consistent agreement.

\subsection{Initial conditions and parameters}

In our simulations, a star approaches a BBH in a parabolic orbit (see Appendix for details). We focus on nearly head-on encounters occurring in-plane—either retrograde ($i=180^\circ$) or prograde ($i=0^\circ$)—to study their impact on the evolution of equal-mass BBHs. We vary the BBHs' initial semi-major axes, setting the minimum value at $a_i \gtrsim 0.02\,{\rm AU}$. This ensures that BBHs are likely to experience interactions with stars before merging via gravitational wave inspiral (see Figure~\ref{fig:boundary}). We choose the upper limit on the BBH orbital semi-major axis to be 0.2~AU to ensure that the BBHs are compact enough to merge shortly following the collision $t_{\rm GW}\lesssim 1\,\rm{Gyr}$, preserving their spin-orbit configuration at the time of merger.
We adopt BH masses of $M_1 = M_2 = [10, 15, 20]\,M_{\odot}$, informed by our recent N-body studies of BH-star collisions in young star clusters \citep[see Fig. 4 of][]{Kiroglu_2024}. For a small subset of models, we fix the BBH mass and $a_i$ and vary the encounter inclination over a range of $0^\circ$ to $180^\circ$. While most of our models assume initially circular BBH orbits, we also explore a few cases with initial eccentricities $e=[0.5,0.9]$.

\subsection{Analysis of Hydrodynamics}

In post-processing, we compute at each time snapshot the mass bound to each of four components: the two individual BHs, the star (if it survives the collision), and the common envelope that forms around the BBH. We employ an iterative procedure similar to that of \citet{Lombardi_2006} and \citet{Kremer_2022}. A necessary condition for a particle to be bound to one of the components is that it has negative specific mechanical energy relative to the center of mass of that component. In particular, the specific mechanical energy of particle $i$ with respective to the center of mass of component $j$ is calculated as
\begin{equation}
\frac{v_{ij}^2}{2}-\frac{GM_j}{d_{ij}} \label{specific_mechanical_energy}
\end{equation}
where $v_{ij}$ is the speed of particle $i$ relative to the center of component $j$, $G$ is Newton's gravitational constant, $M_j$ is the mass of component $j$, and $d_{ij}$ is the distance between the particle and the center of component $j$.  We choose not to include the specific internal energy $u_i$ in the boundness criterion, the preferred approach of \citet{Nandez_2014}.  If the specific mechanical energy is negative for either the BH or the star, the particle is bound to the corresponding component (or possibly to the common envelope instead, as discussed below); if it is negative for more than one component, it is bound to the one for which it is most negative.

The iterative nature of this procedure arises from the need to refine the gravitational potential used in Equation~\ref{specific_mechanical_energy}. Initially, at each snapshot, we assume the masses bound to the components are the same as those from the previous snapshot. We then calculate the specific mechanical energy of each particle and update the set of bound particles for each component. These updated bound masses are then used to refine the gravitational potential and the specific mechanical energy calculations. The process repeats until the set of bound masses converges, ensuring self-consistency between the calculated bound particles and the gravitational potential associated with them.

Unlike in \citet{Kremer_2022}, where each simulation contained only a single BH, the presence of a BBH requires that we consider a common envelope.  There are two ways by which a particle could be counted as part of the common envelope.  First, particles that are not bound to either BH individually but are bound to the center of mass of the binary system are considered part of the common envelope. This binding criterion uses the center of mass position of the binary and the total mass of the two BHs to determine whether a particle has negative specific mechanical energy relative to the center of mass of the BBH. Second, particles may be considered bound to the common envelope if they lie outside the Roche lobe of either BH, but still have negative specific energy with respect to one or both BHs individually. The Roche lobe radius for each BH is calculated using the standard Eggleton approximation \citep{Eggleton_1983}, with the separation between the two BHs taken as the periapsis separation of the BHs; this criterion ensures that the mass bound to each BH remains stable over an orbit. Particles that are not bound to either of the BHs or the common envelope are considered part of the ejecta.

Asymmetric mass loss during the collision imparts a net momentum to the BBH, causing it to drift within the center of mass frame in which the calculation is performed. To quantify this kick, we calculate the center of mass velocity of the BBH, including the mass bound to it, at the final simulation snapshot, defining this velocity as $v_{\rm kick}$.

We note that the calculation of the orbital parameters by the post-processing code treats the BHs, including the mass bound to them, in the two-body approximation. This simplification
occasionally leads to an assignment of a BBH eccentricity $e>1$ during the disruption of the star (see Figure~\ref{fig:evol_insp}).
This temporary artifact arises when the relative velocity of the BHs exceeds the escape velocity calculated in a two-body framework, that is, calculated without regard for the existence of the star (or common envelope). In three-body interactions, the star can boost the BHs’ relative speed,  resulting in values of eccentricity $e>1$ being calculated even though the BBHs will ultimately remain as a binary. The eccentricity can then drop below 1 as the star or stellar debris repositions itself and decreases the relative speed of the two BHs.  The eccentricity values being shown are most reliable once the star has been fully disrupted and the system is well modeled as two bodies.

\subsection{Mass Accretion and Spin-up}

In order to compute $\chi_{\rm eff}$ (Equation~\ref{eq:chi_eff}) for each BBH, we must define how the spin vector of each BH evolves following this accretion. We calculate the spin attained by an initially non-rotating BH with mass $M$ due to accretion of the disrupted material following \citet{Thorne_1974,Bardeen_1972}:
\begin{equation}
    \chi(M,M_{\rm f}) = \left(\frac{2}{3}\right)^{1/2} \frac{M}{M_{\rm f}}\left[4-\left(\frac{18M^2}{M_{\rm f}^2}-2\right)^{1/2}\right]
\end{equation}
where $M_{\rm f}$ is the final BH mass after accreting.
This prescription assumes that the specific angular momentum accreted is at most the angular momentum of the last stable orbit. The final BH mass is derived from our hydrodynamic simulations, under the assumption that all material determined to be bound to each BH is eventually accreted. Our estimates therefore represent an upper limit to the BH accretion and spin values. We assume the direction of each BH spin is in the same direction as the final angular momentum of the disk.

Although in reality accretion and spin-up occur on longer timescales than our SPH simulations, it is also useful to define an ``instantaneous'' $\chi_{\rm eff}$ value, calculated from the angular momentum of stellar debris bound to each BH at any given time. This parameter evolves as both the amount of material and its orientation change over time, and helps inform how the angular momentum is exchanged between the BBH and the gas over the course of the encounter. This is distinct from the \textit{final} effective spin parameter $\chi_{\rm eff,f}$  (reported in Column 18 of Table \ref{table:collision_outcomes}) measured at the end of our simulations, once the system's evolution has stabilized.

\section{Results of hydrodynamic models }
\label{sec:results}
In this section, we describe the results of hydrodynamic simulations of BBH-star collisions and explore their effects on the BBH orbital parameters and the BH spins, both in direction and magnitude. We provide a list of all hydrodynamic simulations performed in this study in Table \ref{table:collision_outcomes}, including the initial conditions and outcomes.
\begin{figure}
    \centering
\includegraphics[width=1.\linewidth]{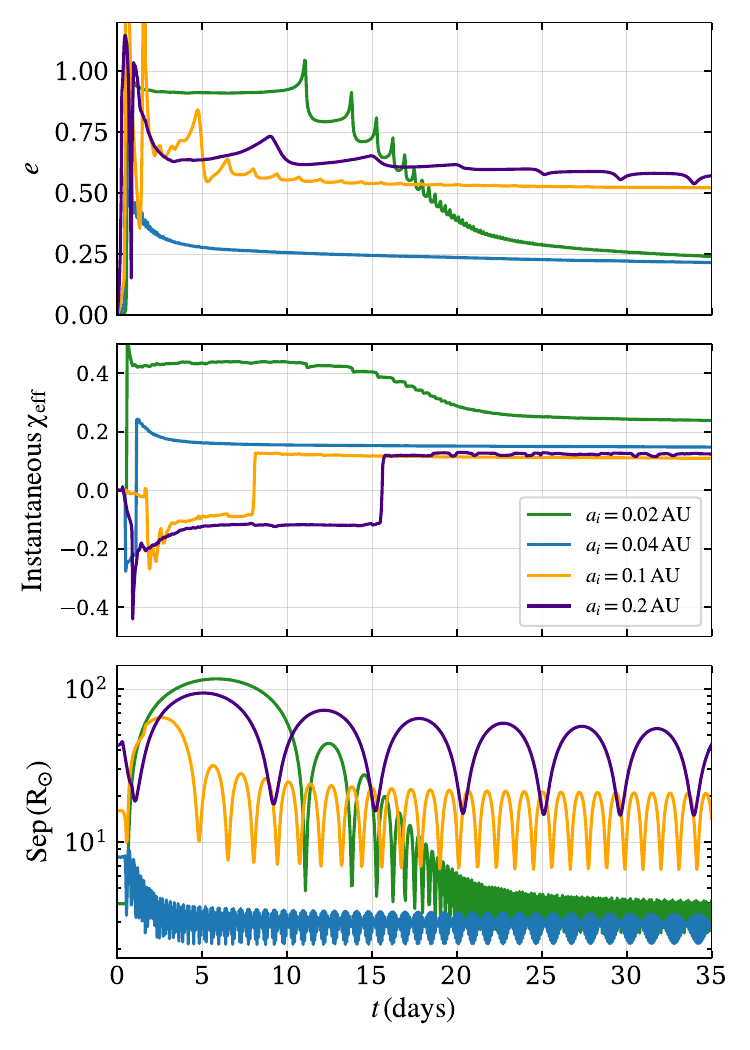}
        \caption{\footnotesize From top to bottom, time evolution of the eccentricity, instantaneous effective spin parameter ($\chi_{\rm eff}$), and separation between BBHs after their collision with a $10\,M_{\odot}$ star. Models with initial BBH semi-major axes ranging from 0.02 to 0.2 AU are shown in different colors (cases 1, 16, 23 and 38 in Table \ref{table:collision_outcomes}, respectively). For cases with $a \geq 0.1\,\rm{AU}$, $\chi_{\rm eff}$ starts negative and transitions to positive during the closest approach between the BHs. This transition is consistently accompanied by a decrease in the semi-major axis, driven by angular momentum transfer from the BBH to the stellar debris. In contrast, for $a < 0.1\,\rm{AU}$, the instantaneous $\chi_{\rm eff}$ is initially positive and remains so throughout the evolution. As the stellar debris approaches steady state, the BBH orbit continues to shrink and circularize, ejecting a significant fraction of the mass in the common envelope around the orbit (see Column 12 in Table~\ref{table:collision_outcomes}).}
        \label{fig:evol_insp}
\end{figure}
\begin{figure*}
    \centering
    \begin{minipage}[t]{0.29\textwidth}
        \centering
        \begin{overpic}[trim=0 0 192 0,clip,width=\linewidth]{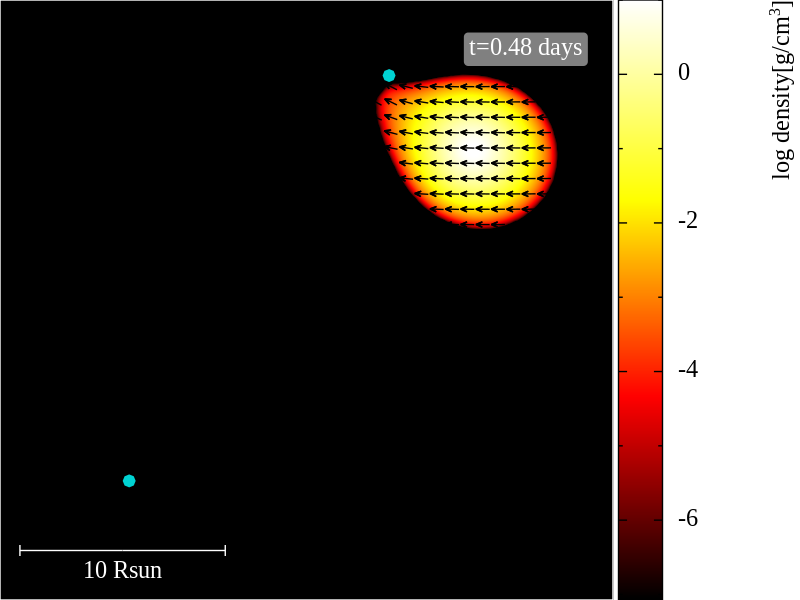}
         \put(5, 89){\color{white}(a)} 
         \put(60, 11){\color{white}$\chi_{\rm eff}=0$}
        \end{overpic}
    \end{minipage}
    \hfill
    \begin{minipage}[t]{0.29\textwidth}
        \centering
        \begin{overpic}[trim=0 0 192 0,clip,width=\linewidth]{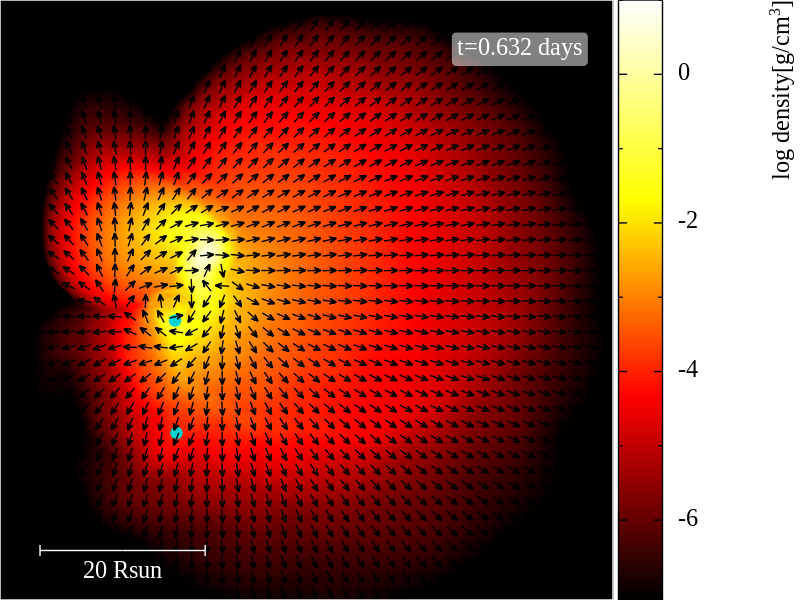}
        \put(5, 89){\color{white}(b)}
        \put(60, 11){\color{white}$\chi_{\rm eff}=-0.07$}
        \end{overpic}
    \end{minipage}
    \hfill
    \begin{minipage}[t]{0.3817\textwidth}
        \centering
        \begin{overpic}[width=\linewidth]{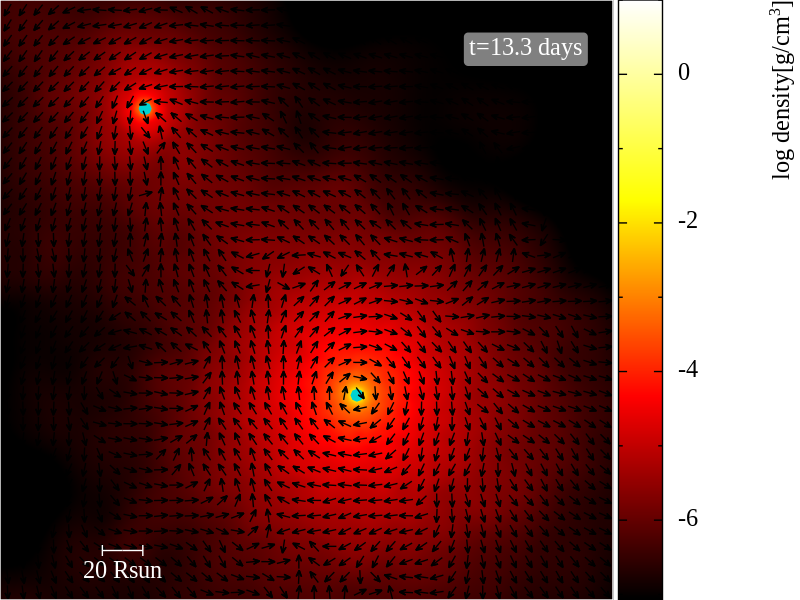}
         \put(4.1, 67.8){\color{white}(c)}
         \put(45, 8.45){\color{white}$\chi_{\rm eff}=-0.29$}
        \end{overpic}
   \end{minipage}
    
    \vspace{1em}
    \begin{minipage}[t]{0.29\textwidth}
        \centering
        \begin{overpic}[trim=0 0 192 0,clip,width=\linewidth]{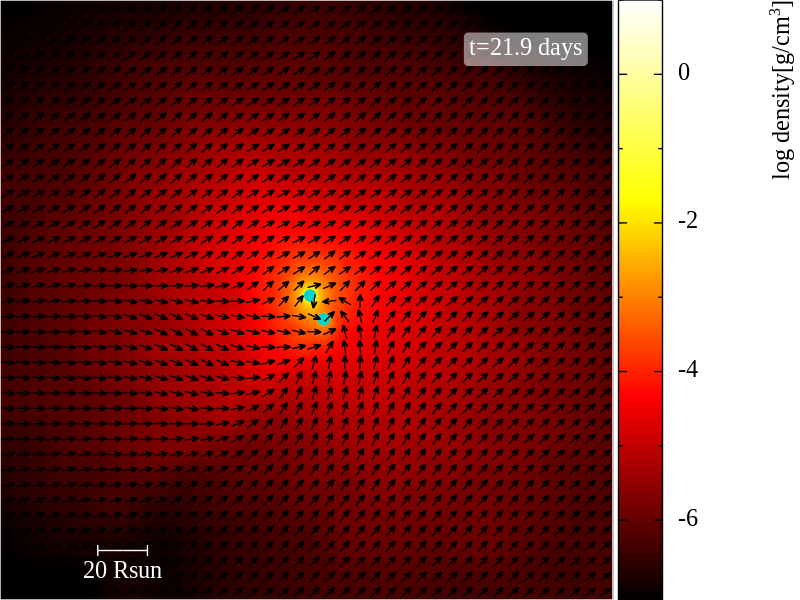}
        \put(5, 89){\color{white}(d)}
        \put(60, 11){\color{white}$\chi_{\rm eff}=-0.28$}
        \end{overpic}
    \end{minipage}
    \hfill
    \begin{minipage}[t]{0.29\textwidth}
        \centering
        \begin{overpic}[trim=0 0 192 0,clip,width=\linewidth]{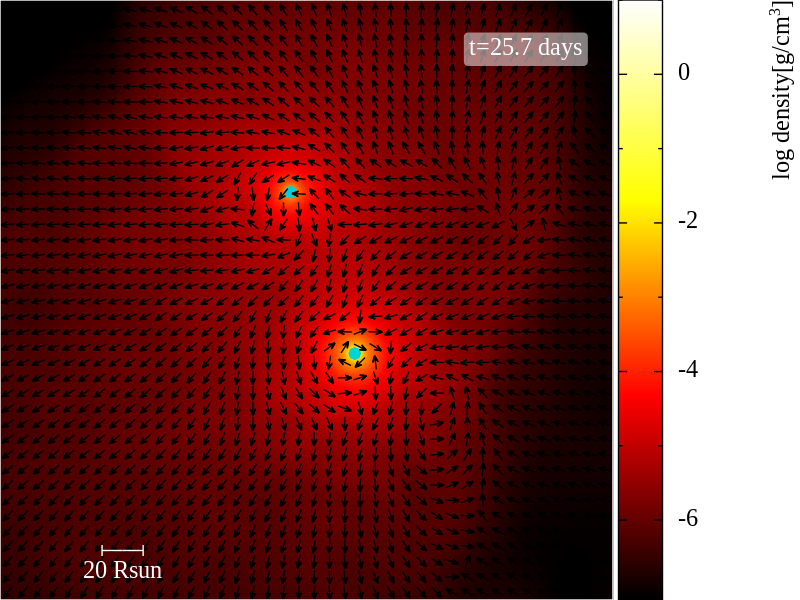}
         \put(5, 89){\color{white}(e)}
         \put(60, 11){\color{white}$\chi_{\rm eff}=-0.19$}
        \end{overpic}
   \end{minipage}
    \hfill
    \begin{minipage}[t]{0.3817\textwidth}
        \centering
        \begin{overpic}[width=\linewidth]{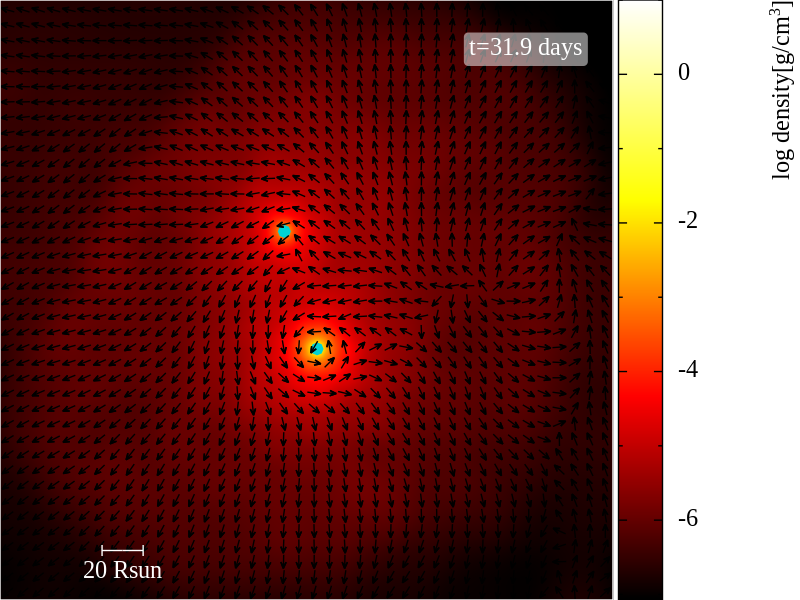}
         \put(45, 8.45){\color{white}$\chi_{\rm eff}=+0.29$}
        \end{overpic}
    \end{minipage}
    
    \caption{\footnotesize Density cross-section snapshots at progressively later times in the 10 MS case (model 9 in Table~\ref{table:collision_outcomes}).
    The BHs, represented by cyan data points, are orbiting counterclockwise.
    Arrows show the local direction of the velocity field in the frame of BH\,2, which is the upper BH in frames (a), (b), and (d), and the lower BH in the remaining frames.
    The star is beginning to be disrupted by BH\,2 in frame (a), and leaves, as can be seen in frame (b), a disk orbiting clockwise (resulting in an instantaneous $\chi_{\rm eff} < 0$). The BHs go through a periapsis passage, and then are near apoapsis in frame (c). At this point, BH\,2 (near bottom) has the more substantial disk, and it orbits clockwise (so  the instantaneous $\chi_{\rm eff}$ is still negative). Frame (d) is a periapsis passage that begins to reverse the disk direction. In frame (e), near apoapsis, the disk around BH\,2 (near bottom) has had
    some
    of its rotation removed. By Frame (f), the next apoapsis, the disk around BH\,2 (near bottom) has completely reversed, resulting in final $\chi_{\rm eff} > 0$. An animation of this figure is available at this \href{https://drive.google.com/file/d/15Fo7HOF4eHIlYxjm2iu4gBhFJNEdnGf8/view?usp=share_link}{link}. 
    }
    \label{fig:density_snapshots}
\end{figure*}
\begin{figure}
    \centering
\includegraphics[width=1.\linewidth]{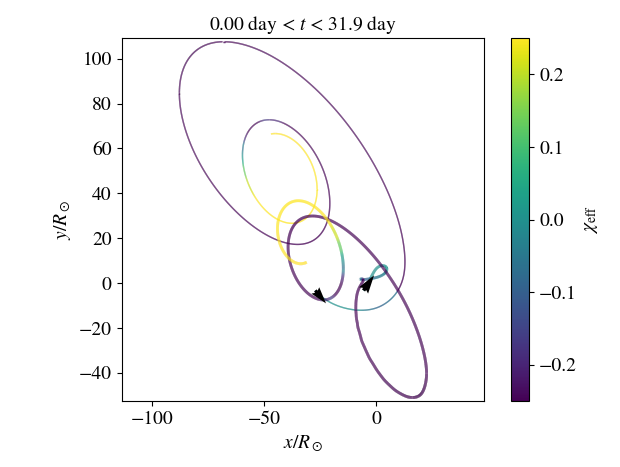}
        \caption{\footnotesize Trajectories of the two black holes for the same case shown in Fig.\ \ref{fig:density_snapshots}, with colors representing the instantaneous $\chi_{\rm eff}$. The triangular arrows show the direction of motion at $t=0$.  The thin curve denotes BH1 and the thick curve denotes BH2.  Notice how the $\chi_{\rm eff}$ value changes as the BHs approach and go through a periapsis passage, due to angular momentum exchange from the orbit to the disks.}
    \label{fig:trajectory}
\end{figure}

\begin{figure}[t]
    \centering
\includegraphics[width=1\linewidth]{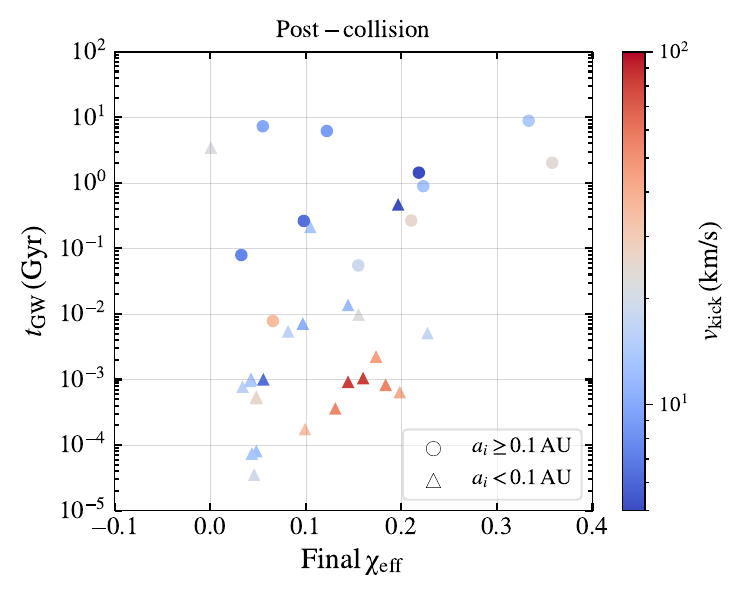}
    \caption{\footnotesize The final inspiral timescale versus final effective spin of BBHs after colliding with a $10\,M_{\odot}$ star. In all cases, the BBH achieves a positive $\chi_{\rm eff}$ value as a consequence of angular momentum exchange between the BBH orbit and stellar debris. Initially compact BBHs ($a_i <0.1\,\rm{AU}$; represented by triangles) generally attain relatively short post-collision inspiral times ($\lesssim 10\,$Myr) and therefore are expected to merge before subsequent dynamical interactions can randomize their positive spin-orbit alignment. For initially wide BBHs ($a_i \geq\,0.1 \rm{AU}$; represented by circles), the post-collision spin-orbit alignment may ultimately be randomized by subsequent encounters within their host cluster, unless the binaries are ejected from their host following the collision. The color of each point indicates the final center-of-mass velocity of the BBH, which arises from linear momentum kicks imparted by ejected stellar material.}
        \label{fig:t_gw_chi_eff}
\end{figure}

In the region of the parameter space we consider, where $a_i \lesssim 0.2\,\rm{AU}$, each BH in a binary acquires a rotating envelope following the disruption of the star. This outcome can occur under two distinct scenarios.
In the first case, the star is fully disrupted by one BH, and a subfraction of the stellar debris is captured and bound to the non-disrupting BH. This outcome is equivalent to the ``overflow scenario'' described in \citet{Lopez_2019}.
The second case involves multiple pericenter passages, where the star survives the first encounter with one BH and subsequently undergoes additional pericenter passage(s) around both BHs, ultimately leading to its complete disruption. 
For very wide BBH orbits, i.e., $a_i \gtrsim 1\,\rm{AU}$, the collision is expected to resemble a single BH-star interaction with only the disrupting BH accreting material \citep{Lopez_2019}. In this regime (where $a_i \gg R_{\star}$), binary torques cannot influence the angular momentum of the disks, which is determined by the angular momentum of the star at the time of disruption. Consequently, these cases are not of interest for this study.

In Figure \ref{fig:evol_insp}, we present the evolution of the eccentricity, instantaneous $\chi_{\rm eff}$, and separation between the BHs following their collision with a $10\,M_{\odot}$ star. The four colors show separate simulations with different initial semi-major axes. For the binaries with $a_i \geq 0.1\,\rm{AU}$, the instantaneous $\chi_{\rm eff}$ is negative after the initial disruption of the star. We quantify this phase with the parameter $t_{\chi <0}$ (shown in column 17 in Table \ref{table:collision_outcomes}), which is the longest consecutive time interval that the instantaneous $\chi_{\rm eff}$ is less than zero. As the BBH undergoes subsequent pericenter passages, the binary exerts a torque on the misaligned stellar debris and ultimately tilts its orientation into alignment with the BBH orbit. In Column 16 of Table \ref{table:collision_outcomes}, we list the time (typically around $1-10\,$days) between the disruption of the star and the next pericenter passage of the BHs. As tidal torques that lead to angular momentum exchange are most prominent at the pericenter, this timescale is typically comparable to the timescale for the angular momentum of the stellar debris to align with the orbit.

In Figure~\ref{fig:density_snapshots}, we show a series of SPH time snapshots for the $a_i =0.1\,\rm{AU}$ model of Figure~\ref{fig:evol_insp} (model 13 in Table~\ref{table:collision_outcomes}). Here, color denotes the mass density of stellar material and arrows denote velocity vectors of the fluid elements in the plane shown. As notated in the various panels, the instantaneous $\chi_{\rm eff}$ is initially negative but ultimately flips to a positive value as angular momentum exchanges from the BBH orbit to the gas. In Figure \ref{fig:trajectory}, we show trajectories for the two BHs in this same simulation, with colors representing the instantaneous $\chi_{\rm eff}$. Again, we see that after multiple pericenter passages, each providing a significant tidal torque, the instantaneous $\chi_{\rm eff}$ ultimately flips from negative to positive.

Conversely, for the binary in Figure~\ref{fig:evol_insp} with an initial semi-major axis of $a_i = 0.02\,\rm{AU}$, the stellar debris is aligned with the orbit after the initial disruption of the star. In this case the instantaneous $\chi_{\rm eff}$ is positive at all times ($t_{\chi < 0} = 0$ in Table~\ref{table:collision_outcomes}).
These collective findings show that spin-orbit alignment is ultimately achieved in \underline{all} BBH-star collisions involving stars and BBHs with the mass ratios and orbital properties explored in this work, where the BBH orbit is comparable to the characteristic size of the envelope, roughly $R_\star$.

 In most of the simulations performed (nearly head-on encounters), the initial angular momentum of the star's orbit is less than that of the BBH orbit. Consequently, the stellar debris field will tend to align with the orbital direction of the BBH. However, in cases where the angular momentum in the star's orbit is large (grazing encounters; (models 28 and 31 in Table \ref{table:collision_outcomes}), angular momentum exchange occurs in the opposite direction: the orbital direction of the BBH will tend to align with the debris field. This follows a general principle that the angular momentum of the orbit and the debris field ultimately tend toward alignment, regardless of which component contributes more significantly.

In Figure~\ref{fig:t_gw_chi_eff}, we present the final inspiral timescale versus the final effective spin $\chi_{\rm eff,f}$ of the post-collision BBHs for all SPH simulations 
In response to the stellar collision, we find BBH orbits can decrease by up to 10 times (the ratio of final to initial semi-major axis, $a_f/a_i$, is shown in Column~14 of Table~\ref{table:collision_outcomes}) and attain moderate eccentricity $e\approx 0.5$. Both of these factors reduce their gravitational wave inspiral timescales significantly. As a result, many of the post-collision BBHs shown in Figure~\ref{fig:t_gw_chi_eff} are likely to merge with positive $\chi_{\rm eff}$, before subsequent dynamical encounters can randomize their spin-orbit orientation.

A fraction of the material from the disrupted star becomes bound to the BHs, while a fraction, with positive total energy, becomes unbound from the system entirely. This unbound ejecta imparts a recoil kick to the center of mass of the BBH. As shown in Column 12 of Table \ref{table:collision_outcomes}, the amount of unbound material can be comparable to the BH mass.

The cases with high kick velocities are associated with encounters that have a large impact velocity and a direct, close to head-on collision between the star and one of the BHs. Such cases result in asymmetric mass ejection that gives a significant kick to the BBH.  In the context of a star colliding with a single BH, this kick is studied in some detail in \citet[][see Section 3.2]{Kremer_2022}. The main difference here is that the kick is ultimately given to the BBH as a whole instead of a single, isolated BH.  In cases when the impact velocity is small or the encounter with the BH is less direct, much of the mass ejection occurs during the subsequent common envelope-like phase of the BBH, leading to more axisymmetric ejection and a smaller final kick.

The resulting recoil velocities can reach up to $80\,$km/s, potentially sufficient to eject the BBH from its host cluster core or, in extreme cases, from its host cluster entirely. In these cases, even the BBHs with relatively long post-collision inspiral times ($t_{\rm GW} \gtrsim 100\,$Myr) may still retain positive $\chi_{\rm eff}$ values at merger, since the chance of a subsequent encounter is reduced in the lower-density cluster halo (or eliminated fully in the case of ejection from the cluster). The kick velocities of BBHs after each collision are illustrated in the color bar of Fig. \ref{fig:t_gw_chi_eff}.

\section{Conclusion and Discussion}
\label{sec:conc}
BBHs formed either dynamically or through binary evolution in dense star clusters are expected to undergo a series of interactions, including collisions and mergers with stars. These interactions can significantly alter their mass, spin, and orbital properties, thereby shaping their GW signatures during inspiral and merger. To investigate these processes, we performed a suite of SPH simulations of close encounters between massive stars and BBHs typical of those expected to occur in young and dense cluster environments.

Previous studies have explored the impact of stellar tidal disruption events on BBH spin-orbit orientations \citep{Lopez_2019,Ryu_2022}. However, these studies have focused mainly on the disruption of $1\,M_{\odot}$ main-sequence stars, motivated by both the higher abundance of low-mass stars in clusters and the shorter lifetimes of more massive stars. In this work, we explore interactions with massive ($10\,M_{\odot}$) stars that lead to direct collisions with the BBH components. We obtain qualitatively different results, as the impact on the binary orbit and mass accretion is greater in nearly-equal mass encounters. 
We find that collisions between BBHs and massive stars result in the preferential alignment of the BH component spins with the orbital angular momentum, provided the binaries are initially sufficiently compact ($\lesssim 1\,\rm{AU}$) to form individual accretion disks that can interact with and be torqued by the orbit during the close pericenter passage. On the other hand, \citet{Ryu_2022} found that initially circular BH binaries experience only $ \approx 10\% $ increase in their eccentricity following the disruption of a $1\,M_{\odot}$ star. In that case, the periapsis passage may never get close enough to induce significant torques for the disk to realign with the orbital angular momentum.

Our $N$-body cluster simulations demonstrate a typical cluster features approximately 10 (3) BBH+star collisions with mass ratios $q>0.1$ ($q>0.5$) at early times (see Fig.~\ref{fig:boundary}). Consequently, for a typical cluster with roughly 100 BBH mergers over its full $\sim12\,$Gyr lifetime \citep[e.g.,][]{Kremer2020_catalog}, we expect roughly $10\%$ of BBH mergers to be affected by these collisions. Following angular momentum exchange and ultimately accretion, we find these collisions tend to result in spin-orbit alignment of the BBH, with important implications for the BBH's effective spin parameter $\chi_{\rm eff}$ at the time of merger. 
While canonical gas-free cluster dynamics provides the best route for forming binaries with a significant component of the spins anti-aligned with the orbital angular momentum \citep[e.g.,][]{Payne2024}, BBHs undergoing an accretion phase after colliding with massive stars in young clusters could contribute toward the observed bias toward small positive $\chi_{\rm eff}$ values seen in current LVK data. 
The fraction of BH mergers with aligned spins due to interactions with massive stars, as predicted in this study ($\approx 10\%$), is roughly comparable to the fraction of mergers that are asymmetric around $\chi_{\rm eff}=0$ as predicted by \citet{Banagiri_2025}.

A key limitation of the calculations presented here is the exclusion of accretion feedback effects on the hydrodynamics and ultimate accretion. Accretion releases energy over various timescales and may drive outflows, resulting in an uncertain fraction of mass loss. While our hydrodynamic models provide insight into the amount of stellar material bound to each BH, we assume $100\%$ accretion efficiency of this material in our spin calculations, ignoring the potential impact of these outflows. Consequently, our estimates represent an upper limit to the BH spin values. Additionally, although our simulations are hydrodynamic, in reality magnetic fields are present, which may drive accretion either via turbulence that acts like an effective viscosity or via winds that extract mass and angular momentum via torque.

For cases where the initial disks are misaligned with the orbit, we assume that the disks persist for at least ~10 days, long enough for the next BH periapsis passage to occur. However, it is possible that the material in the misaligned disk might have already been accreted by the BH. In this case, the disk could remain misaligned long enough for the BH to accrete the anti-aligned angular momentum, possibly leading to a final negative effective spin parameter. 
The viscous accretion timescale remains highly uncertain as it depends on the poorly constrained alpha viscosity and disk scale height. Further simulations may provide further insight into these quantities.

Last, if the common envelope around the BBH formed after the collision survives until the time of the BBH merger, it could produce an EM counterpart, resembling a weak short gamma-ray burst \citep{Perna_2016}. A detailed exploration of such counterparts associated with accreting BBH mergers in dense star clusters will be presented in a forthcoming paper (Kıroğlu et al. 2025, in preparation).

\begin{acknowledgements}

This work was supported by NSF grant AST-2108624 and NASA grant 80NSSC22K0722, as well as the computational resources and staff contributions provided for the \texttt{Quest} high-performance computing facility at Northwestern University. \texttt{Quest} is jointly supported by the Office of the Provost, the Office for Research, and Northwestern University Information Technology. F.K.\ acknowledges support from a CIERA Board of Visitors Graduate Fellowship.
We thank Emma Chambers and Cade Snyder for their contributions to earlier related work that aided the development of analysis routines used in this research.
This work made use of the SPLASH visualization software \citep{2007PASA...24..159P}.

\end{acknowledgements}

\appendix
\subsection{Initial Conditions for Black Hole and Star Positions and Velocities}

To facilitate comparisons with future works, we describe here the initial conditions for the positions and velocities of the BHs and the star. 
The overall strategy uses the analytic solution to the Kepler two-body problem \citep[e.g.,][]{thornton2004classical} twice. The first usage of the Keplerian solution is for the BBH: we treat the BHs as a two-body system orbiting their common center of mass in an $XY$ plane. The orbit of the BBH is then rotated out of that plane at an inclination angle $i$, as described below. The second usage of the Keplerian solution is for the parabolic orbits of the star and the BBH about their common center of mass. To keep this center of mass of the three-body system at the origin, we shift the BBH and star in opposite directions away from the origin of $xyz$ frame used for the SPH calculation.

To initiate the BBH, we set values for the following parameters: the masses of the two BHs ($M_{\rm BH1}=M_{\rm BH2}=M_{\rm BBH}/2$), the semi-major axis $a$,
the orbital eccentricity $e$,
the inclination $i$ relative to the orbital plane of the star, and the true anomaly $f$. In all of our simulations, the initial argument of periapsis and longitude of the ascending node are zero. An $XY$ coordinate system is used to define, temporarily, the BBH orbital dynamics.
In our convention, a true anomaly $f=0$ corresponds to the second BH being on the $+X$-axis.
Because we consider equal mass BHs, the position coordinates of the first BH in the $XY$ frame are simply $X_1 = -\frac{R}{2} \cos f$ and $Y_1 = -\frac{R}{2} \sin f$, where the separation of the BH components is
\begin{equation}
R=\frac{a(1-e^2)}{1 + e \cos f}. \label{separation}
\end{equation}
The position of the second BH is then chosen so that the center of mass of the binary is at the origin in the $XY$ frame: $X_2=-X_1$, $Y_2=-Y_1$.
The velocity components of the first BH in this orbital plane are obtained by differentiating $X_1$ and $Y_1$ with respect to time, treating the semimajor axis $a$ and eccentricity $e$ as constants, with
\begin{equation}
\dot{f} = (1+e\cos f)^2\sqrt{\frac{G M_{\rm BBH}}{a^3(1-e^2)^3}}
\end{equation}
as the initial rate of change of the true anomaly
and
\begin{equation}
\dot{R} = e\sin f\sqrt{\frac{G M_{\rm BBH}}{a(1-e^2)}} \label{Rdot}
\end{equation}
as the initial rate of change of the separation.
The velocity components of the second BH are similarly set, but with the opposite sign so that the total momentum of the BBH in the $XY$ frame is zero.

For the initial conditions of the star, the BBH is treated as a single point mass at its center of mass, and both the BBH and the star follow parabolic orbits about their common center of mass.
Their separation is given by $r=2r_p/(1-\cos \theta)$, where the angular position $\theta$ of the star is measured counterclockwise with respect to the $+x$-axis in the $xy$ plane and the periapsis separation $r_{\rm p}$ occurs at $\theta=\pi$.

The parabolic trajectory of the star is aligned such that it approaches the BBH from infinity in the first quadrant and would recede toward infinity in the fourth quadrant.  (When $r_p=0$, the star approaches from along the $+x$-axis.)

To set the initial position of the star, we determine its coordinates for the initial separation $r=r_0=50 R_\odot$ and the initial angular position $\theta=\theta_0 = \cos^{-1}(1- 2r_p/r_0)$.
In particular, so that the center of mass of the entire three-body problem is at the origin in the $xyz$ frame, the star is placed with position coordinates
    $x_3 = (1+Q)^{-1} r \cos \theta$, 
    $y_3 = (1+Q)^{-1} r \sin \theta$, and
    $z_3=0$,
while the initial center of mass position of the BBH is $x_{\rm BBH}=-Qx_3$, $y_{\rm BBH}=-Q y_3$, $z_{\rm BBH}=0$. Here the mass ratio $Q=M_3/M_{\rm BBH}$ and $M_3$ is the star mass.  We note that $x_3-x_{\rm BBH}=r \cos\theta$ and $y_3-y_{\rm BBH}=r \sin\theta$.

The initial velocity components of the star are obtained by differentiating $x_3$ and $y_3$ with respect to time and making use of
    $\dot{\theta} = \sqrt{2G r_p M_{\rm tot}}/r^{2}$ and
    $\dot{r} = -\sqrt{2 G M_{\rm tot}(1-r_p/r)/r}$ (from angular momentum and energy conservation), where $M_{\rm tot}=M_{\rm BBH} + M_3$.
The center of mass velocity of the BBH is then chosen so that the center of mass of the entire three-body system will remain at the origin in the $xyz$ frame.
 
The 3D positions of the binary components, accounting for the center of mass shift and the inclination $i$, are 
\begin{equation}
\begin{bmatrix}
x_j \\ y_j \\ z_j
\end{bmatrix}
=
\begin{bmatrix}
x_{\rm BBH} \\ y_{\rm BBH} \\ 0
\end{bmatrix}+
\begin{bmatrix}
\cos i & 0 & -\sin i \\
0 & 1 & 0 \\
\sin i & 0 & \cos i
\end{bmatrix}
\begin{bmatrix}
X_j \\ Y_j \\ 0
\end{bmatrix},
\end{equation}
where \(j = 1, 2\) corresponds to the two BHs, and \(X_j, Y_j\) are the coordinates of the BHs in the orbital plane, with \(Z_j = 0\).
The transformation of the velocity components is of the same form as the transformation of the position coordinates. We note that $i=0$ corresponds to a prograde approach, while $i=180^\circ$ corresponds to a retrograde approach.


Putting everything together, the initial position of the star is
\begin{equation}
\begin{bmatrix}
x_3 \\ y_3 \\ z_3
\end{bmatrix}
= (1+Q)^{-1} 
\begin{bmatrix}
   r_0 - 2r_p\\
   2 ( r_0 r_p- r_p^2)^{1/2} \\
   0
\end{bmatrix},
\end{equation}
while the initial positions of the BHs are
\begin{equation}
\begin{bmatrix}
x_j \\ y_j \\ z_j
\end{bmatrix}
=
\begin{bmatrix}
-Qx_3\mp\frac{R}{2}\cos i \cos f \\ -Qy_3\mp\frac{R}{2}\sin f \\ \mp\frac{R}{2}\sin i \cos f
\end{bmatrix},
\end{equation}
where $R$ is given by Equation (\ref{separation}) and where the minus sign is chosen from the $\mp$ for BH
$j=1$ and the plus sign is chosen for BH $j=2$.
The initial velocity of the star is
\begin{equation}
\begin{bmatrix}
v_{x3} \\ v_{y3} \\ v_{z3}
\end{bmatrix}
=-(1+Q)^{-1}\sqrt{2 G M_{\rm tot}}
\begin{bmatrix}
    \sqrt{\left(1-r_p/r_0\right)/r_0}\\
   \sqrt{r_p}/r_0 \\
   0
\end{bmatrix}
\end{equation}
while the initial velocities of the BHs are
\begin{equation}
\begin{bmatrix}
v_{xj} \\ v_{yj} \\ v_{zj}
\end{bmatrix}
=
\begin{bmatrix}
   -Qv_{x3}\pm\frac{1}{2}(R\dot{f} \sin f -\dot{R}\cos f) \cos i \\
   -Qv_{y3}\mp\frac{1}{2}(R\dot{f}\cos f + \dot{R}\sin f) \\
   \pm\frac{1}{2}(R\dot{f} \sin f -\dot{R}\cos f)\sin i
\end{bmatrix},
\end{equation}
where $R$, $\dot{f}$, and $\dot{R}$ come from Equations (\ref{separation})-(\ref{Rdot}) and 
where the top sign is chosen from the $\mp$ or the $\pm$ for BH $j=1$ and the bottom sign is chosen for BH $j=2$.
For example, in Case 3, 
we have in code units ($G=M_\odot=R_\odot=1$) $Q=0.5$, $a=3.95$, $e=0$, $f=30^\circ$, $r_p=0$, $i=0$, $M_{\rm BBH}=20$, which yields initial positions for the star and BHs of (33.3, 0.0) and ($-16.7 \mp 1.7$, $\mp 1.0$), respectively, as well as initial velocities of (-0.73, 0) and ($0.37\pm 0.56$, $\mp 0.97$).
In Case 25,
we have in code units $Q=1/3$, $a=21.5$, $e=0.5$, $f=0^\circ$, $r_p=0.5$,
$i=180^\circ$, $M_{\rm BBH}=30$,
which yields initial positions for the star and BHs of (36.75, 7.46) and
($-12.250 \mp 5.375$, -2.488),
respectively, as well as initial velocities of (-0.944, -0.095) and
(0.315, 0.032$\mp 1.023$).

\begin{figure}
    \centering
    \includegraphics[width=0.8\linewidth]{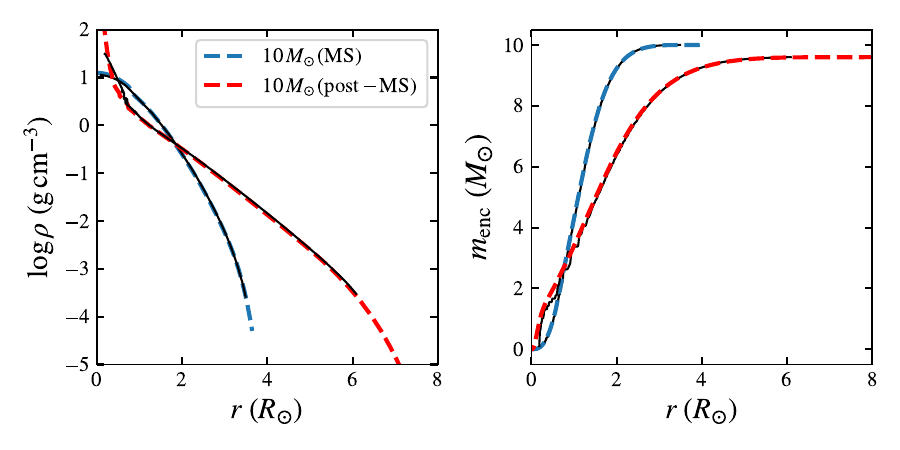}
    \caption{\footnotesize The density and mass profiles of the MESA stellar models (dashed lines) in comparison to their SPH models at the end of relaxation (solid lines).  }
        \label{fig:mesa}
\end{figure}

\newpage

\startlongtable
\begin{rotatetable*}
\begin{deluxetable*}{cccccccccccccccccccc}
\label{table:collision_outcomes}
\tabletypesize{\tiny}
\tablewidth{600pt}
\tablecaption{Collision Outcomes and Post-Collision Properties}
\tablehead{
    \colhead{{$^1$}Case} & 
    \colhead{{$^2$}$M_{\rm BBH}$} & 
    \colhead{{$^3$}star} & 
    \colhead{{$^4$}$r_p$} & 
    \colhead{{$^5$}$a_{\rm i}$} & 
    \colhead{{$^6$}$e_{\rm BBH}$} & 
    \colhead{{$^7$}$f$} & 
    \colhead{{$^8$}$i$} & 
    \colhead{{$^9$}$t_{\rm GW, i}$} & 
    \colhead{$^{10}$$M_{\rm BH1,f}$} & 
    \colhead{$^{11}$$M_{\rm BH2,f}$} & 
    \colhead{$^{12}$$m_{\rm ej}$} & 
    \colhead{$^{13}$$m_{\rm CE}$} & 
    \colhead{$^{14}$$a_{\rm f}/a_{\rm i}$} & 
    \colhead{$^{15}$$e_{\rm f}$} & 
    \colhead{$^{16}$$t_{\rm dtp}$} &
    \colhead{$^{17}$$t_{\chi < 0}$} & 
    \colhead{$^{18}$$\chi_{\rm eff,f}$} & 
    \colhead{$^{19}$$t_{\rm GW, f}$} &
    \colhead{$^{20}$ $v_{\rm kick}$} \\
    &
    $[M_\odot]$ & 
 & 
    [AU] & 
    [AU] & 
    &
    [deg] &
    [deg] & 
    [Myr] & 
    $[M_\odot]$ & 
    $[M_\odot]$ & 
    $[M_\odot]$ & 
    $[M_\odot]$ & 
    & 
    & 
    [day] &
    [day] &
    &
    [Myr] &
    [km/s]
}
\startdata
1 & 20 & 10\,MS & 0.012 & 0.018 & 0.0 & 0 & 0 & 18 & 11.25 & 10.21 & 7.8 & 0.77 & 0.78 & 0.12 & 11 & 0 & 0.23 & 5.1 & 17 \\ 
2 & 20 & 10\,MS & 0.000 & 0.018 & 0.0 & 0 & 0 & 18 & 10.20 & 10.07 & 8.9 & 0.84 & 0.33 & 0.48 & 0.47 & 0.64 & 0.05 & 0.081 & 13 \\ 
3 & 20 & 10\,MS & 0.000 & 0.018 & 0.0 & 30 & 0 & 18 & 10.18 & 10.08 & 8.4 & 1.29 & 0.27 & 0.48 & 0.24 & 0.35 & 0.05 & 0.035 & 20 \\ 
4 & 20 & 10\,MS & 0.000 & 0.018 & 0.0 & 0 & 30 & 18 & 10.01 & 10.00 & 9.6 & 0.43 & 0.21 & 0.73 & 0.3 & 0.33 & 0.00 & 0.0022 & 30 \\ 
5 & 20 & 10\,MS & 0.000 & 0.018 & 0.0 & 0 & 180 & 18 & 10.07 & 10.18 & 9.0 & 0.75 & 0.32 & 0.49 & 0.46 & 0.62 & 0.04 & 0.074 & 14 \\ 
6 & 20 & 10\,MS & 0.007 & 0.018 & 0.0 & 0 & 0 & 18 & 10.95 & 10.28 & 8.1 & 0.66 & 0.45 & 0.01 & 1.6 & 0.0092 & 0.20 & 0.64 & 41 \\ 
7 & 20 & 10\,MS & 0.000 & 0.037 & 0.0 & 0 & 0 & 2.9e+02 & 10.19 & 10.59 & 8.5 & 0.72 & 0.19 & 0.05 & 0.36 & 0 & 0.13 & 0.36 & 54 \\ 
8 & 20 & 10\,MS & 0.000 & 0.037 & 0.0 & 0 & 180 & 2.9e+02 & 10.95 & 10.19 & 5.6 & 3.24 & 0.26 & 0.31 & 0.083 & 0 & 0.18 & 0.83 & 55 \\ 
9 & 20 & 10\,MS & 0.002 & 0.100 & 0.0 & 0 & 0 & 1.6e+04 & 10.07 & 11.40 & 6.6 & 1.97 & 0.66 & 0.50 & 21 & 27 & 0.22 & 8.9e+02 & 13 \\ 
10 & 20 & 10\,MS & 0.000 & 0.100 & 0.0 & 30 & 0 & 1.6e+04 & 10.86 & 10.20 & 8.4 & 0.52 & 0.12 & 0.17 & 8.2 & 8.3 & 0.17 & 2.2 & 45 \\ 
11 & 20 & 10\,MS & 0.000 & 0.100 & 0.0 & 60 & 0 & 1.6e+04 & 10.05 & 11.22 & 7.1 & 1.66 & 0.67 & 0.63 & 19 & 25 & 0.20 & 4.7e+02 & 5.1 \\ 
12 & 20 & 10\,MS & 0.005 & 0.100 & 0.0 & 0 & 0 & 1.6e+04 & 10.05 & 12.44 & 5.0 & 2.52 & 1.5 & 0.65 & 88 & 89 & 0.33 & 8.9e+03 & 14 \\ 
13 & 20 & 10\,MS & 0.009 & 0.100 & 0.0 & 0 & 0 & 1.6e+04 & 10.01 & 12.79 & 4.8 & 2.38 & 0.75 & 0.37 & 23 & 31 & 0.36 & 2e+03 & 24 \\ 
14 & 20 & 10\,MS & 0.000 & 0.200 & 0.0 & 0 & 0 & 2.6e+05 & 10.01 & 11.46 & 4.4 & 4.13 & 0.36 & 0.46 & 37 & 39 & 0.22 & 1.4e+03 & 4.1 \\ 
15 & 30 & 10\,MS & 0.000 & 0.018 & 0.0 & 0 & 0 & 5.4 & 15.48 & 15.38 & 6.8 & 2.35 & 0.47 & 0.28 & 0.17 & 0.16 & 0.10 & 0.17 & 34 \\ 
16 & 30 & 10\,MS & 0.000 & 0.037 & 0.0 & 0 & 0 & 91 & 15.21 & 16.09 & 8.0 & 0.71 & 0.34 & 0.18 & 0.38 & 0.67 & 0.14 & 0.92 & 82 \\ 
17 & 30 & 10\,MS & 0.000 & 0.037 & 0.0 & 0 & 30 & 91 & 15.15 & 15.25 & 9.1 & 0.53 & 0.34 & 0.46 & 0.35 & 0.055 & 0.05 & 0.52 & 23 \\ 
18 & 30 & 10\,MS & 0.000 & 0.037 & 0.0 & 0 & 60 & 91 & 15.26 & 15.10 & 9.0 & 0.61 & 0.47 & 0.58 & 0.26 & 0.32 & 0.04 & 1 & 12 \\ 
19 & 30 & 10\,MS & 0.000 & 0.037 & 0.0 & 0 & 90 & 91 & 15.14 & 15.15 & 9.0 & 0.75 & 0.36 & 0.41 & 0.3 & 0.14 & 0.03 & 0.78 & 16 \\ 
20 & 30 & 10\,MS & 0.000 & 0.037 & 0.0 & 0 & 120 & 91 & 15.10 & 15.25 & 9.0 & 0.65 & 0.47 & 0.59 & 0.25 & 0.32 & 0.04 & 0.95 & 14 \\ 
21 & 30 & 10\,MS & 0.000 & 0.037 & 0.0 & 0 & 150 & 91 & 15.26 & 15.15 & 9.0 & 0.55 & 0.35 & 0.47 & 0.36 & 0.065 & 0.05 & 0.55 & 26 \\ 
22 & 30 & 10\,MS & 0.000 & 0.037 & 0.0 & 0 & 180 & 91 & 16.26 & 15.21 & 7.7 & 0.86 & 0.36 & 0.26 & 0.39 & 0.68 & 0.16 & 1 & 82 \\ 
23 & 30 & 10\,MS & 0.000 & 0.074 & 0.0 & 0 & 0 & 1.5e+03 & 15.10 & 15.84 & 7.6 & 1.45 & 0.83 & 0.52 & 3.2 & 6.4 & 0.10 & 2.1e+02 & 13 \\ 
24 & 30 & 10\,MS & 0.002 & 0.100 & 0.0 & 0 & 180 & 4.8e+03 & 15.00 & 15.45 & 8.9 & 0.70 & 0.15 & 0.38 & 7.5 & 0.68 & 0.05 & 1.3 & 17 \\ 
25 & 30 & 10\,MS & 0.002 & 0.100 & 0.5 & 0 & 180 & 1.7e+03 & 15.13 & 15.27 & 9.0 & 0.58 & 0.2 & 0.67 & 0.26 & 0.014 & 0.05 & 0.94 & 6.5 \\ 
26 & 30 & 10\,MS & 0.002 & 0.100 & 0.9 & 0 & 180 & 16 & 15.00 & 15.00 & 10.0 & 0.04 & 0.042 & 0.99 & 0.59 & 0.54 & 0.00 & 1.1e-08 & 65 \\ 
27 & 30 & 10\,MS & 0.000 & 0.100 & 0.0 & 0 & 180 & 4.8e+03 & 15.08 & 15.61 & 8.5 & 0.81 & 0.76 & 0.82 & 8.3 & 3.8 & 0.08 & 31 & 8.7 \\ 
28 & 30 & 10\,MS & 0.047 & 0.100 & 0.0 & 0 & 180 & 4.8e+03 & 15.00 & 15.00 & 10.0 & 0.03 & 0.037 & 0.85 & 0.6 & 0.6 & 0.00 & 0.00012 & 27 \\ 
29 & 30 & 10\,MS & 0.012 & 0.100 & 0.0 & 0 & 180 & 4.8e+03 & 15.04 & 15.00 & 9.9 & 0.07 & 0.075 & 0.79 & 0.16 & 0.0092 & 0.00 & 0.005 & 17 \\ 
30 & 30 & 10\,MS & 0.012 & 0.100 & 0.5 & 0 & 180 & 1.7e+03 & 15.00 & 15.00 & 9.9 & 0.05 & 0.057 & 0.87 & 0.36 & 0.35 & 0.00 & 0.00038 & 30 \\ 
31 & 30 & 10\,MS & 0.012 & 0.100 & 0.9 & 0 & 180 & 16 & 15.00 & 15.06 & 9.9 & 0.02 & 0.06 & 0.81 & 15 & 0.18 & 0.01 & 0.0016 & 51 \\ 
32 & 20 & 10\,post & 0.000 & 0.040 & 0.0 & 0 & 0 & 4.1e+02 & 10.92 & 10.08 & 8.3 & 0.30 & 0.41 & 0.05 & 0.99 & 1.6 & 0.16 & 9.8 & 22 \\ 
33 & 20 & 10\,post & 0.000 & 0.100 & 0.0 & 0 & 0 & 1.6e+04 & 10.06 & 11.31 & 7.2 & 1.00 & 0.41 & 0.28 & 1.7 & 3.9 & 0.21 & 2.7e+02 & 25 \\ 
34 & 20 & 10\,post & 0.000 & 0.150 & 0.0 & 0 & 0 & 8.1e+04 & 10.09 & 10.90 & 7.5 & 1.11 & 0.23 & 0.54 & 10 & 0.77 & 0.15 & 55 & 18 \\ 
35 & 30 & 10\,post & 0.000 & 0.040 & 0.0 & 0 & 0 & 1.2e+02 & 15.66 & 15.18 & 8.1 & 0.65 & 0.55 & 0.30 & 0.52 & 0.65 & 0.10 & 7.1 & 11 \\ 
36 & 30 & 10\,post & 0.000 & 0.100 & 0.0 & 0 & 0 & 4.8e+03 & 15.60 & 15.05 & 8.5 & 0.40 & 0.36 & 0.69 & 1.7 & 1.1 & 0.07 & 7.8 & 35 \\ 
37 & 30 & 10\,post & 0.000 & 0.150 & 0.0 & 0 & 0 & 2.4e+04 & 15.84 & 15.07 & 6.6 & 2.05 & 0.54 & 0.66 & 4.6 & 0.9 & 0.10 & 2.6e+02 & 6.5 \\ 
38 & 30 & 10\,post & 0.000 & 0.200 & 0.0 & 0 & 0 & 7.6e+04 & 15.05 & 16.07 & 6.5 & 1.94 & 0.76 & 0.56 & 8.7 & 15 & 0.12 & 6.2e+03 & 9 \\ 
39 & 40 & 10\,post & 0.000 & 0.040 & 0.0 & 0 & 0 & 51 & 20.51 & 20.51 & 8.0 & 0.56 & 0.62 & 0.27 & 0.4 & 0.092 & 0.08 & 5.4 & 16 \\ 
40 & 40 & 10\,post & 0.000 & 0.100 & 0.0 & 0 & 0 & 2e+03 & 20.48 & 20.10 & 7.6 & 1.46 & 0.77 & 0.68 & 0.36 & 0.75 & 0.03 & 79 & 7.5 \\ 
41 & 40 & 10\,post & 0.000 & 0.200 & 0.0 & 0 & 0 & 3.2e+04 & 20.11 & 20.68 & 7.9 & 0.90 & 0.95 & 0.54 & 9.1 & 9.3 & 0.06 & 7.3e+03 & 9.9 \\ 
\hline
\enddata
\normalsize
\tablecomments{Columns 1–8 list the initial conditions for each simulation, including the total BBH mass, stellar model, pericenter distance, BBH semi-major axis, BBH eccentricity, BBH true anomaly, and BBH inclination. Columns 9 and 19 report the inspiral timescale of the BBH before and after the collision, respectively. Columns 10–15 describe the final outcomes, including the final masses of the BHs, the mass of ejected material, the envelope mass surrounding the BBH, the ratio of the final to initial BBH semi-major axis, and the final BBH eccentricity. Column 16 lists the amount of time between the disruption of the star to pericenter passage of the BHs. We define "disruption" to be when the mass bound to the more disruptive BH reaches $10\%$ the peak bound mass to that BH.  
Column 17 lists the amount of time the stellar debris is misaligned with the orbit (instantaneous $\chi_{\rm eff} > 0$) before becoming aligned (instantaneous $\chi_{\rm eff} > 0$). Column 18 lists the final effective spin parameter, calculated using Eq. \ref{eq:chi_eff}. Column 20 lists the kick velocity the center of the BBH receives due to mass ejection form the system.  }
\end{deluxetable*}
\end{rotatetable*}

\bibliography{mybib}{}
\bibliographystyle{aasjournal}

\end{document}